\documentclass[11pt,a4paper]{article}
\usepackage{amsmath}
\usepackage{jheppub}
\usepackage{graphicx,graphics}
\usepackage{dcolumn} 
\usepackage{bm} 
\usepackage{amssymb}
\usepackage[caption=false,labelformat=simple]{subfig}

\newcommand{\gsim}{\gtrsim}
\begin{document}
%%%%%%%%%%%%%%%%%%  Title %%%%%%%%%%%%%%%%%%%%%%%%%%%%%%%%%%%%%%%%%%%%
\title{Novel signals for the Type-X two Higgs doublet scenario at the
Large Hadron Collider}

%%%%%%%%%%%%%%%%%%  Authors  %%%%%%%%%%%%%%%%%%%%%%%%%%%%%%%%%%%%%%%%%
\author[a]{Biswarup Mukhopadhyaya}
\emailAdd{biswarup@iiserkol.ac.in}

\author[a]{Sirshendu Samanta}
\emailAdd{ss21rs027@iiserkol.ac.in}

\author[a]{Tousik Samui}
\emailAdd{tousiksamui@gmail.com}

\author[a]{Ritesh K. Singh}
\emailAdd{ritesh.singh@iiserkol.ac.in}
\affiliation[a]{Department of Physical Sciences, Indian Institute of
Science Education and Research Kolkata, Mohanpur, 741246, India.}
%%%%%%%%%%%%%%%%%%%%%%%%%%%%%%%%%%%%%%%%%%%%%%%%%%%%%%%%%%%%%%%%%%%%%%

\abstract{
We consider, in the context of the Large Hadron Collider, the signals
of the Type-X two Higgs doublet model (2HDM) in the parameter region
answering the best possible solution to the muon $(g-2)$ data within
this framework. The analysis
takes into account all theoretical and observational constraints, and is based
on the final state comprising a same-sign dilepton pair and a pair of
same-sign $\tau$ jets. The crucial ingredient in making the signal clean is the
same-sign feature of both the dilepton and the $\tau$-jet pair individually. After
a detailed estimate of the signal and all
noteworthy backgrounds, we show that this channel offers by far the best
signal significance among those studied so far, predicting discovery with
an integrated luminosity of 3000~fb$^{-1}$, and strong indications
even with 1000~fb$^{-1}$ if systematic uncertainies do not exceed about 10\%.
We also demonstrate that the recently developed dynamic radius jet algorithm
is effective in this connection.
}

\maketitle

%%%%%%%%%%%%%%%%%%%%%%%%%%%%%%%%%%%%%%%%%%%%%%%%%%%%%%%%%%%%%%%%%%%%%%
\section{Introduction}\label{sec:intro}
Whether more than one scalar $SU(2)$ doublets are responsible for the
still enigmatic phenomenon of electroweak symmetry breaking (EWSB) continues
to remain an open issue. One reason behind this is the repetitive
occurrence of spin-1/2 fields, and the unanswered query as to why
scalar fields in the electroweak (EW) theory should be immune to such
repetition. Thus there remains continued interest in two Higgs doublet
models (2HDM)\,\cite{Branco:2011iw,Wang:2022yhm}, the simplest
examples of an extended scalar sector. Such scenarios are consistent
with electroweak precision tests (EWPT), and, in spite of the data on the
125-GeV scalar suggesting closeness to the
`alignment limit'\,\cite{Bhattacharyya:2015nca}, the potential for new
phenomenology is quite rich. Such prospects, however, depend on what
type of 2HDM it is. Since the unconstrained coupling of both the
doublets to  $T_3 = +1/2$ as well as $-1/2$ fermions can lead to
tree-level flavour violation, a frequent practice is to impose
$\mathbb{Z}_2$ symmetries on the Yukawa terms in various ways, so that
each fermion couples to one doublet only. One thus ends up with models
belonging to Type-I, Type-II, Type-X or `Lepton-Specific' and 'Flipped'
type\,\cite{Branco:2011iw,Wang:2022yhm}. The experimental signatures,
too, depend on which type one is concerned with. We are concerned with
some signals of the Type-X 2HDM in this paper.

Type-X 2HDM is a scenario where one of the two scalar doublets (called
$\Phi_2$) couples to quarks only, and the other one (called $\Phi_1$),
to leptons alone. While a review of the model can be found in the
literature\,\cite{Jueid:2021avn}, and is summarised in
section~\ref{sec:model} as well, we mention here a few salient
features which affect its associated phenomenology. First of all, the
scalar states dominated by $\Phi_1$ have large branching ratios to
dilepton final
states\,\cite{Kanemura:2011kx,Kanemura:2014bqa}. Secondly, the neutral
pseudoscalar ($A$) can be rather light ($\gsim 62$ GeV), consistent
with all current limits \,\cite{CMS:2018qvj}. And thirdly, in certain
regions of the parameter space, it implies substantial new
contributions to the muon anomalous magnetic
moment\,\cite{Dey:2021pyn,Muong-2:2021ojo,Muong-2:2021vma}.
Although some recent estimates (largely based on observed rates for
$h\to 4\tau$\,\cite{ATLAS:2018emt,ATLAS:2018pvw,CMS:2018qvj,CMS:2018nsh,CMS:2018zvv,CMS:2019spf}) disfavour $m_A \le m_h/2$\,\cite{Dey:2021pyn}, this scenario is still
consistent with the $(g-2)_\mu$ measurement within 3$\sigma$ level
over a wide range of parameter space. Therefore, this model deserves a
serious attention in this context, especially in view of the
persistent uncertainty in theoretical predictions of $(g-2)_\mu$.

It is thus important to closely examine the predictable signals of
Type-X 2HDM at the Large Hadron Collider (LHC), particularly when the
high-luminosity run takes place \,\cite{Tomas:2022mzg}. Though some
earlier studies have partially constrained the parameter space\,\cite{Jueid:2021avn},
one needs to proactively devise search strategies using various final
states pertaining to this particular model, where a considerable scope
for improvement remains is still there.

Studies have taken place, suggesting reconstruction of $A$ in the
$\mu^+\mu^-$ channel\footnote{It should be noted
that the best signal significance in this channel was predicted for
$m_A < m_h/2$, which has subsequently been disfavored by the data in
$h\to 4\tau$\,\cite{ATLAS:2018emt,Dey:2021pyn,ATLAS:2018pvw,CMS:2018qvj,CMS:2018nsh,CMS:2018zvv,CMS:2019spf}.}\,\cite{Chun:2017yob}. It is also important to
find signatures of the heavier neutral and charged scalars
$H,H^{\pm}$. For this, the muonic channel of at least one $A$ which
results from decays of the heavier states have been made use
of\,\cite{Chun:2018vsn}. Although this results in apparently clean
signals, rates are suppressed  by the $A \rightarrow \mu^+ \mu^-$
branching ratio, thus making it difficult to rise above the
$3\sigma$-level with an integrated luminosity of $3000$~fb$^{-1}$.

We go beyond these studies and consider instead  $4\tau$ signals which
arise, for example, via the hard scattering channels
$pp \rightarrow H A, H^\pm A$, followed by $H \rightarrow ZA$
and $H^\pm \rightarrow W^\pm A$. Each of the two $A$'s thus produced
decays dominantly to a $\tau$ pair. The novelty of our approach lies
in the following points:
\begin{itemize}
\item The events corresponding to charged and neutral heavy scalars
    can be clubbed together since their masses are constrained to be
    small from electroweak precision observables. Moreover, we analyse
    events where the $Z$ or the $W$ decays into jets. 

\item Out of the $4\tau$ final state, we have concentrated on events
    where two same-sign $\tau$'s have one-and three-prong hadronic decays,
    while the remaining $\tau$-pair, also of the same sign, decay
    leptonically. In order to do so, we have utilised the claim that
    the $\tau$-induced jets can have charge identification efficiencies
    of 99\%  and 70\% in the one-and three-prong channels,
    respectively\,\cite{CMS:2022prd}. Thus one looks for a pair of
    same-sign leptons as well as a pair of same-sign tau-jets. After
    convolution with the appropriate tau-identification efficiencies,
    and on using suitable event selection criteria, one thus ends up
    with substantial signal rates along with a rather impressive
    background reduction.

\item In addition to the decays $H \rightarrow ZA$, we have included
    cases where the $H$ directly decays into a $\tau$-pair, thus
    yielding events similar to those mentioned above. The additional
    jets arise from showering. This inclusion boosts the strength of
    the signal.

\item We have used a recently developed dynamic radius jet
    algorithm\,\cite{Mukhopadhyaya:2023rsb} which is demonstrated to
    be as good as the anti-$k_t$ algorithm. 
\end{itemize}

The paper is organised as follows. A brief outline of the Type-X 2HDM
has been provided in section \ref{sec:model}, together with the
existing constraints on the parameter space. The choice of benchmarks
for our analysis is thus motivated. Section \ref{sec:sig-bkg}
contains a full-length discussion of the proposed signal and its
various backgrounds, which leads to the adopted  event selection
strategy. The results are presented and discussed in section
\ref{sec:res}. We summarise and conclude in section \ref{sec:summ}.

%%%%%%%%%%%%%%%%%%%%%%%%%%%%%%%%%%%%%%%%%%%%%%%%%%%%%%%%%%%%%%%%%%%%%%
\section{Type-X 2HDM: parameters and constraints} \label{sec:model}
As has been already mentioned, Type-X 2HDM envisions a situation
where, in the Higgs flavour basis, $\Phi_2$ has Yukawa interactions
with all quarks, and $\Phi_1$, with leptons. This is ensured by
imposing a $\mathbb{Z}_2$ symmetry  on the Yukawa interaction, under
which the fields transform as
\begin{eqnarray}
&\Phi_1 \rightarrow - \Phi_1; &\qquad \Phi_2 \rightarrow \Phi_2; \\
&Q_L, Q_R, L_L \rightarrow Q_L,Q_R,L_L; &\qquad  L_R \rightarrow -L_R,
\end{eqnarray}
where the subscripts $L, R$ stand for left and right-chiral
projections, respectively.

The scalar potential, neglecting CP-violation, is given by
\begin{eqnarray}
V_\text{scalar}
&=&  m_{11}^2 \Phi_1^\dagger \Phi_1
    + m_{22}^2 \Phi_2^\dagger \Phi_2
    + \lambda_1 \left(\Phi_1^\dagger \Phi_1\right)^2
    + \lambda_2 \left(\Phi_2^\dagger \Phi_2\right)^2 \nonumber \\ 
& & + \lambda_3 \left(\Phi_1^\dagger \Phi_1\right)
     \left(\Phi_2^\dagger \Phi_2\right)
    + \lambda_4 \left(\Phi_1^\dagger \Phi_2\right)
     \left(\Phi_2^\dagger \Phi_1\right) \nonumber\\
& & + \left\{ -m_{12}^2 \Phi_1^\dagger \Phi_2
    + \frac{\lambda_5}{2} \left(\Phi_1^\dagger \Phi_2\right)^2
    + \text{h.c.} \right\}, \label{eqn:LagS}
\end{eqnarray}
It should be noted that the $\mathbb{Z}_2$ is broken above by the soft term
proportional to $m^2_{12}$ which does not re-introduce flavour
violation at the tree level.
After the spontaneous EWSB, the two Higgs doublets acquire vacuum expectation values
(vevs) $v_1$ and $v_2$, which usually are reparametrize as
$v= \sqrt{v^2_1 + v^2_2}$ and $\tan\beta = v_2/v_1$. 
In terms of these parameters and the neutral scalar mixing angle
$\alpha$, it is possible to express the physical masses of the spin-0
particles, namely $h,H$ (the neutral scalars), $A$ (the neutral
pseudoscalar) and $H^\pm$ (the charged scalars):
\begin{eqnarray}
      m_{H}^{2}~
&=&    M^2\,s_{\alpha-\beta}^2
    + \left(\lambda_1\,c^2_\alpha\,c^2_\beta
    + \lambda_2\,s^2_\alpha\,s^2_\beta\
    + \frac{\lambda_{345}}{2}\,s_{2\alpha}\,s_{2\beta}\right)v^2,\\
      m_h^2~\,
&=&   M^2\,s^2_{\alpha-\beta}
    + \left(\lambda_1\,s^2_\alpha\,c^2_\beta
    + \lambda_2\,c^2_\alpha\,s^2_\beta\
    - \frac{\lambda_{345}}{2} s_{2\alpha}\,s_{2\beta}\right)v^2,\\
      m_A^2~\,
&=&   M^2-\lambda_5\,v^2, \\
      m_{H^\pm}^2
&=&   M^2-\frac{\lambda_4+\lambda_5}{2} v^2,
\end{eqnarray}
where $M^2 = m^2_{12}/(s_\beta\,c_\beta)$ and, for an angle $\theta$,
$s_\theta(c_\theta)$ represents $\sin\theta(\cos\theta)$. Finally, once
the scalar, pseudoscalar and charged scalar mass matrices are
diagonalised and the Goldstone bosons are separated out, the Yukawa
interactions of the various mass eigenstates are given by
\begin{eqnarray}
      \mathcal{L}_\text{Yukawa}
&=& - \sum_{f} \frac{m_f}{v} \left(\xi_f^h \bar f f h
    + \xi_f^H \bar f f H - i \xi_f^A \bar f \gamma_5 f A \right)
      \nonumber\\
& & - \frac{\sqrt{2}}{v}\Big[V_{ud}^\text{CKM}
      \left(m_u\xi_u^A \bar{u}_R d_L
    + m_d\xi_d^A \bar{u}_L d_R\right)H^+
    + m_\ell \xi_\ell^A \bar{\nu}_L\ell_R H^+
    + \text{h.c.}\Big]\!,
\end{eqnarray}
The case-by-case details of these couplings for various fermions are
summarised in Table~1.
%=====================================================================
\begin{table}[!h]\renewcommand{\arraystretch}{1.25}
\begin{center}
\begin{tabular}{|c|c|c|c|c|c|}
\hline
  $\xi_u^h=\xi_d^h$ & $\xi_\ell^h$ & $\xi_u^H=\xi_d^H$ & $\xi_\ell^H$
& $\xi_u^A = -\xi_d^A$ & $\xi_\ell^A$ \\
\hline
  $\dfrac{\cos\alpha}{\sin\beta}$ & $-\dfrac{\sin\alpha}{\cos\beta}$
& $\dfrac{\sin\alpha}{\sin\beta}$ & $\dfrac{\cos\alpha}{\cos\beta}$
& $\cot\beta$ & $\tan\beta$ \\
\hline
\end{tabular}
\end{center}
\caption{Scale factors of the SM fermion couplings to the 2HDM
physical scalars.}
\label{tab:xi}
\end{table}
%=====================================================================

For the current analysis, we have implemented the model in the
Mathematica-based package
{\tt SARAH}\,\cite{Staub:2008uz,Staub:2013tta} to generate {\tt Universal FeynRules Output} (UFO)\,\cite{Degrande:2011ua} and
{\tt SPheno} \cite{Porod:2003um,Porod:2011nf} compatible output. The
{\tt SPheno} is then used to generate a spectrum with masses and
couplings for a given input parameter point.

In this study, $h$ is identified as the observed 125~GeV scalar at the
LHC\,\cite{Aad:2012tfa,Chatrchyan:2012ufa}. The other CP-even physical
scalar $H$ is kept heavier than $h$. The well-measured masses of the
gauge bosons, namely $W$ and $Z$ bosons, are controlled by vev ($v$).
This fixes the value of $v$ at 246
GeV\,\cite{ParticleDataGroup:2022pth}. The remaining parameters in the
scalar sector are treated as free, subject to constraints from
theoretical considerations and experimental measurements. The
following constraints are relevant here:
\begin{description}
\item [Theoretical constraints\,] The electroweak symmetry breaking
minimum for the scalar potential corresponds to a stable vacuum,
provided\,\cite{PhysRevD.75.035001,2013JHEP...06..045B}
$$\lambda_{1,2} > 0, \qquad \lambda_3 > -\sqrt{\lambda_1\lambda_2},
\quad \text{and} \quad \lambda_3 + \lambda_4 - |\lambda_5| >
-\sqrt{\lambda_1\lambda_2}\,.$$
Furthermore, we restrict the model to satisfy the perturbative unitary
constraints. All the quartic couplings, therefore, should satisfy
$|\lambda_i| < 4\pi$, ($i=1,2,\cdots,6$), in order for the Lagrangian
to be perturbative. Further, tree-level unitarity in any scalar-scalar
to scalar-scalar scattering demands that the real part of each term in
the partial wave decomposition of $2\to2$ scattering amplitude should
be smaller than 1/2. This leads to the following conditions on the
$\lambda$ parameters\,\cite{Wang:2022yhm,Lee:1977yc,2000hep.ph...12353A}:
\begin{eqnarray}
a_\pm &=& \frac{3}{2} \left(\lambda_1 + \lambda_2\right)
         \pm \sqrt{\frac{9}{4}\left(\lambda_1 - \lambda_2\right)^2
        + \left(2\lambda_3 + \lambda_4\right)^2} \leq 8\pi,\\
b_\pm &=& \frac{1}{2}\left(\lambda_1+\lambda_2\right)
          \pm \sqrt{\frac{1}{4}\left(\lambda_1 - \lambda_2\right)^2
        + \lambda_4^2} \leq 8\pi, \\
c_\pm &=& \frac{1}{2}\left(\lambda_1+\lambda_2\right)
          \pm \sqrt{\frac{1}{4}\left(\lambda_1 - \lambda_2\right)^2
        + \lambda_5^2} \leq 8\pi, \\
e_\pm &=& \lambda_3 + 2\lambda_4 \pm 3\lambda_5 \leq 8\pi, \\
f_\pm &=& \lambda_3 \pm \lambda_4 \leq 8\pi, \\
g_\pm &=& \lambda_3 \pm \lambda_5 \leq 8\pi,
\end{eqnarray}
where $a_\pm, b_\pm, \cdots, g_\pm$ are the eigenvalues of the
scattering amplitude matrices involving all possible $2\to2$
scalar-scalar scattering.

\item [Higgs properties and scalar searches\,] In this model, the
properties of the 125~GeV scalar are bound to deviate from predictions
of the SM. Although the measured values of its couplings are almost
consistent with the SM prediction, there is a small window in the
experimental measurement where new physics can be accommodated. This,
in turn, restricts the parameters of any given model.
Also, searches for the additional scalars yield upper limits on their
production cross section. Constraints thus arising are included in
publicly available packages called
{\tt HiggsSignals}\,\cite{Bechtle:2020uwn} and
{\tt HiggsBounds}\,\cite{Bechtle:2020pkv,Bahl:2022igd}, which restrict
the parameter space of the model in consideration. 

\item [Oblique electroweak parameters\,] The precision measurement of
the electroweak observables has widely been studied at the Large
Electron Positron (LEP) collider. The essence of these lies in the
Peskin–Takeuchi parameters, namely $S$, $T$, and
$U$\,\cite{Peskin:1990zt,Peskin:1991sw}. In Type-X 2HDM, the values
$U$ parameter is known to be very small. In the limit $U=0$, the
current measured values are $S=-0.01\pm0.07$ and $T=0.04\pm0.06$ with
a 92\% correlation between them\,\,\cite{ParticleDataGroup:2022pth}.
We used the covariance matrix in the $S$-$T$ plane to calculate
$\chi^2$ after the calculation of $S$ and $T$ at one-loop using
{\tt SPheno}\,\cite{Porod:2003um,Porod:2011nf} for each parameter
point. The parameter points are then subject to passing the constraint
at the $90\%$ C.L. 
\end{description}

%%%%%%%%%%%%%%%%%%%%%%%%%%%%%%%%%%%%%%%%%%%%%%%%%%%%%%%%%%%%%%%%%%%%%%
\begin{figure}[!h]
\begin{center}
\subfloat[]{\includegraphics[width=0.5\textwidth]{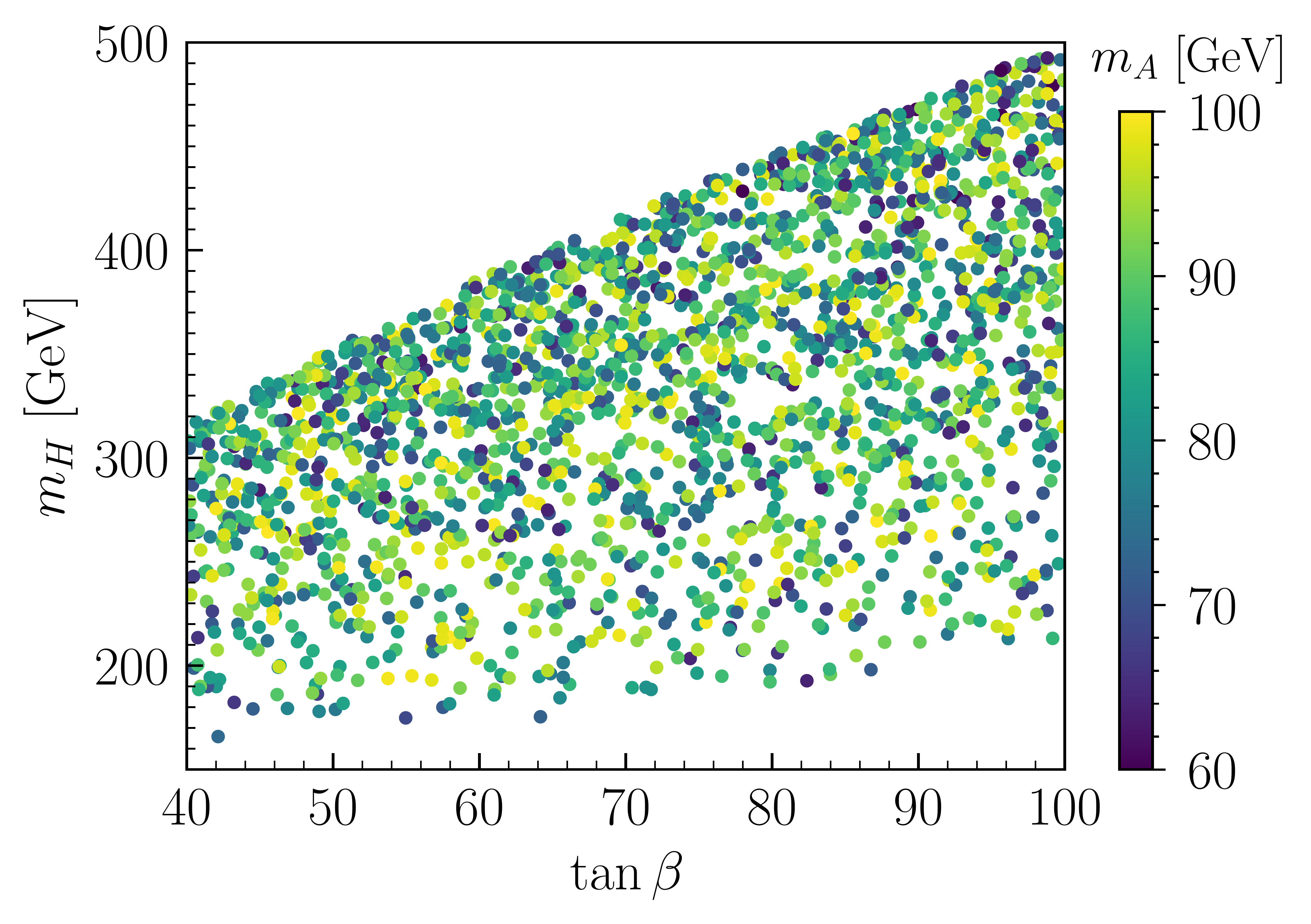}\label{fig:const1}}\hfill
\subfloat[]{\includegraphics[width=0.5\textwidth]{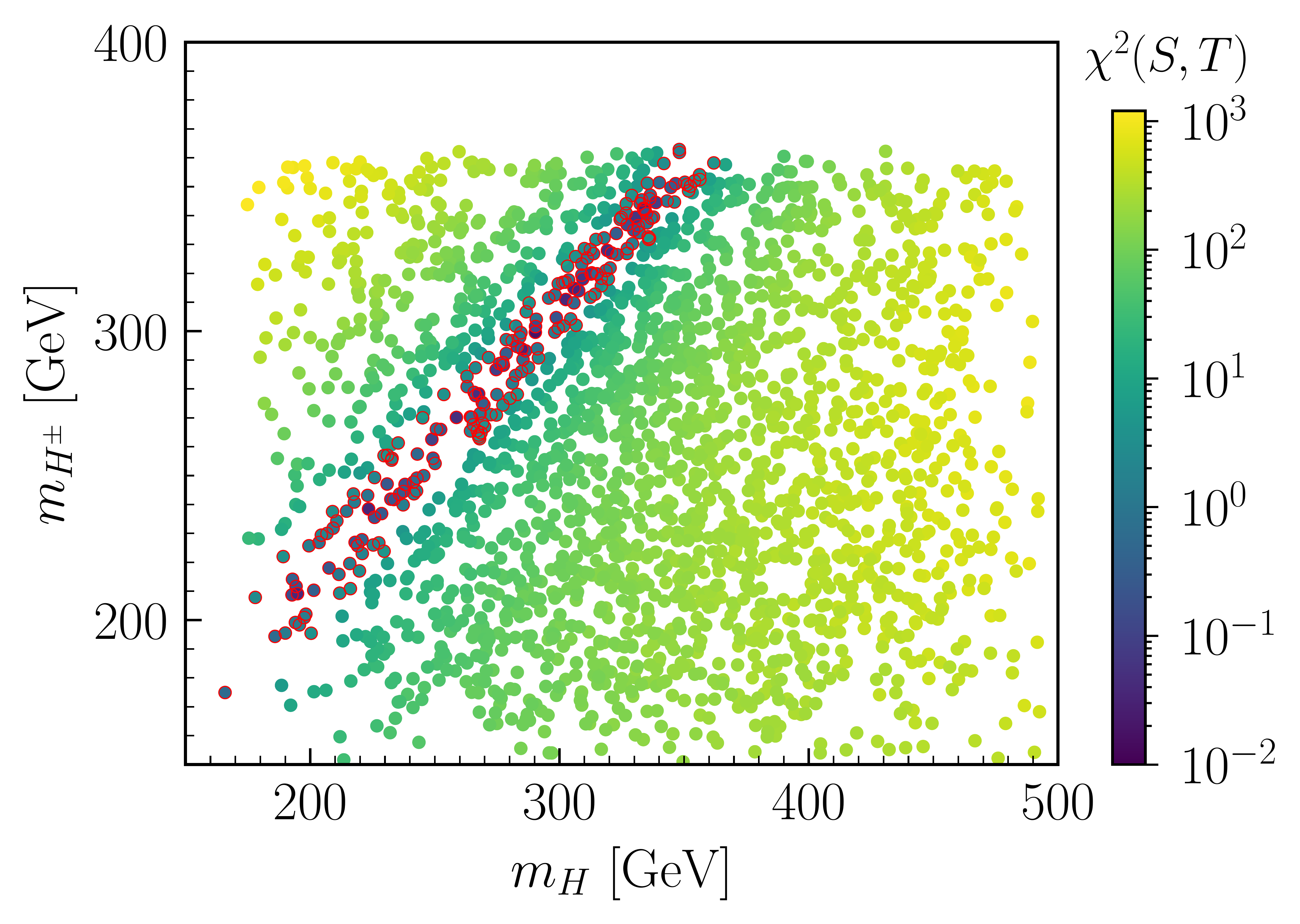}\label{fig:const2}}
\end{center}
% \vspace{-5mm}
\caption[singlelinecheck=on]{(a) Scatter plot of $m_H$ vs.~$\tan\beta$
allowed by {\tt HiggsSignals}\,\cite{Bechtle:2020uwn} and
{\tt HiggsBounds}\,\cite{Bechtle:2020pkv,Bahl:2022igd}. The different
colours on the points represent the variation of the $m_A$. The upper
limits of $m_H$ and $m_{H^\pm}$ are because of the chosen range of
$m_{12}^2$ for the scan. (b) Scatter plot of $m_H$ vs.~$m_{H^\pm}$
allowed by {\tt HiggsSignals} and {\tt HiggsBounds}. The points with
red circles are allowed by the measurement of EW oblique parameter $S$
and $T$ parameter at 95\% C.L.\,\cite{ParticleDataGroup:2022pth}. The
colours of the points represent the value of $\chi^2(S,T)$.}
\label{fig:both}
\end{figure}
% \vspace{-3mm}
%%%%%%%%%%%%%%%%%%%%%%%%%%%%%%%%%%%%%%%%%%%%%%%%%%%%%%%%%%%%%%%%%%%%%%

We have performed a thorough scan of the parameters subject to the
above constraints. For the scan, the six parameters, namely
$\lambda_1$, $m_{12}$, $m_H$, $m_{H^\pm}$, $m_A$, $\tan\beta$, some of
which can be traded off with the quartic couplings in the Lagrangian
given in Eq.~(\ref{eqn:LagS}), have been varied in the following
range:
\begin{eqnarray}
m_H \in (150,500)~\text{GeV}, \quad m_{H^\pm}\in(150,350)~\text{GeV},
\quad m_A\in(60,100)~\text{GeV}, \nonumber\\
\tan\beta\in(40,100), \quad m_{12}^2 \in (450,2500)~\text{GeV}^2,
\quad \text{and} ~~\lambda_1 = 0.1.
\end{eqnarray}

In Figure~\ref{fig:both}, we show the allowed parameter points that
satisfy the above constraints. The points on the $\tan\beta$ and $m_H$
plane in Figure~\ref{fig:const1} are allowed by theoretical and
experimental limits. The feature of having upper limits (approximately
linear in $\tan\beta$) of $m_H$ and $m_{H^\pm}$ is due to the
restriction of the parameter $m_{12}^2$. The lower limit on $m_{12}^2$
is to avoid making various physical masses ($m_A$, $m_{H^\pm}$) too
small to satisfy phenomenological constraints, while the upper limit
confines one to regions where the extended scalar sector lies within
the LHC. The scatter plot in Figure~\ref{fig:const2} shows the allowed
points in the $m_H$-$m_{H^\pm}$ plane. The red circled points are
after the imposition of constraints from the oblique parameters $S$
and $T$. This restricts the mass splitting between $H$ and $H^\pm$ to
small values.

Finally, we take into account the numerical requirements for
explaining $(g-2)_\mu$ which serves as a motivation for Type-X
2HDM\,\cite{Muong-2:2006rrc,Muong-2:2021ojo,Muong-2:2021vma}. Originally, a scenario with $m_A\approx 30$-40~GeV,
consistently with all phenomenology, was found to explain the observed
excess rather nicely\,\cite{Jueid:2021avn}. However, the limit on $4\tau$ searches at
the LHC \,\cite{ATLAS:2018emt,ATLAS:2018pvw,CMS:2018qvj,CMS:2018nsh,CMS:2018zvv,CMS:2019spf} has subsequently brought in some constraints on the on-shell
decay $h\to AA\to 4\tau$. Therefore, it appears more appropriate
if $m_A$ is a little above the pair-production threshold in $h$-decay.
Even after respecting this constraint, one still
finds\,\cite{Dey:2021pyn} a substantial parameter region consistent
with the latest result on
$(g-2)_\mu$\,\cite{Muong-2:2021ojo,Muong-2:2006rrc,Muong-2:2021vma}
at the $3\sigma$ level. The estimate in\,\cite{Dey:2021pyn} includes
contributions from two-loop Bar-Zee diagrams, following
Refs.\,\cite{Queiroz:2014zfa,Ilisie:2015tra}. Our analysis is based on
benchmarks within this region.

Keeping the above discussion in mind, an interesting and at the same
time consistent region in the Type-X 2HDM parameter space is
$$ \frac{m_h}{2} < m_A \lesssim 100~\text{GeV}, \qquad  \tan\beta > 40~\text{GeV},
\qquad 200~\text{GeV} \lesssim m_H \simeq m_{H^\pm} \lesssim 400~\text{GeV}.$$
This prompts our four benchmark points tabulated in
Table~\ref{tab:bp}, for the collider analysis presented in the
following sections. 

%=====================================================================
\begin{table}[!h]\renewcommand\arraystretch{1.25}
\begin{center}
% \begin{tabular}{|p{0.2\textwidth}|p{0.2\textwidth}|p{0.2\textwidth}|p{0.2\textwidth}|}
\begin{tabular}{|c|c|c|c|c|}
\hline
 & \quad $m_A$ (GeV) \qquad & \quad $m_H$ (GeV)\qquad
 & \quad $m_{H^\pm}$ (GeV) \qquad & \quad $\tan\beta$ \qquad \\
%$\textbf{m_A}$ & $\textbf{m_H}$ & $\textbf{m_H^{\pm}}$ & $\textbf{\tan\beta}$ \\
\hline\hline
\quad BP1 \quad\qquad & 63.1  & 210.7 & 204.0 & 61.8 \\
\hline
\quad BP2 \quad\qquad & 63.2 & 249.0 & 250.2 & 60.0 \\
\hline
\quad BP3 \quad\qquad & 70.2 & 217.0 & 213.5 & 69.8 \\
\hline
\end{tabular}
\caption{The set of benchmark points chosen for further collider
studies. All three points are allowed by the theoretical and
experimental constraints described above.}
\label{tab:bp}
\end{center}
\end{table}
%=====================================================================

%%%%%%%%%%%%%%%%%%%%%%%%%%%%%%%%%%%%%%%%%%%%%%%%%%%%%%%%%%%%%%%%%%%%%%
\section{Collider Study}
\subsection{Signal and Background} \label{sec:sig-bkg}
We consider signals arising out of the hard scattering process
$pp\to HA$ and $pp \to H^\pm A$ at the 14~TeV
LHC. The fact that $m_H$ and $m_H^\pm$ are constrained to be closely
spaced enables us to club together these two hard scattering
processes, and analyse the resulting final states with the same
kinematic criteria. The dominant decay modes of $H$ or $H^\pm$ are to
a massive weak boson $Z$ or $W^\pm$ whereas the pseudoscalar $A$
predominantly decays to a pair of $\tau^\pm$ leptons. The weak bosons
decay hadronically. Therefore, the following signals ensue from both of
the above production channels.
\begin{eqnarray}
&pp \to H A \to Z A A \to 4 \tau + \text{jets}\\
&pp \to H^{\pm} A \to W^\pm A A \to 4 \tau + \text{jets}
\end{eqnarray}
In this way, we finally have $2\tau^+ + 2\tau^- + 2j$ as final states
after all the decays cascading from heavy scalars or light
pseudoscalars. Being unstable, the $\tau$ decays to the other
two light leptons with a branching ratio at $\approx$\,35\%.
However, it has a higher branching ratio of $\approx$\,65\%, to jets via
hadronic decay modes. These jets formed out of the hadronic decays of
$\tau$ leptons are usually distinct from light QCD jets due to
their low multiplicity in terms of their constituents and therefore
can be tagged as $\tau$ jets. These jets, usually written as $\tau_h$,
have almost 60\% tagging efficiency with a very small
$\left(\approx\,0.5\%\right)$ mistagging rate\,\cite{CMS:2018jrd} defined as the
fraction at which the other jets, falsely, are being tagged as
$\tau_h$. Even with this relatively high efficiency and really small
mistagging rate, the signal in the said channel will tend to be
swamped by the QCD background, especially in the regions of the
$\tau$-jets having $p_T$ around $10$-$50$~GeV.

\begin{figure}[!h]
\begin{center}
\subfloat[]{\includegraphics[width=0.62\textwidth]{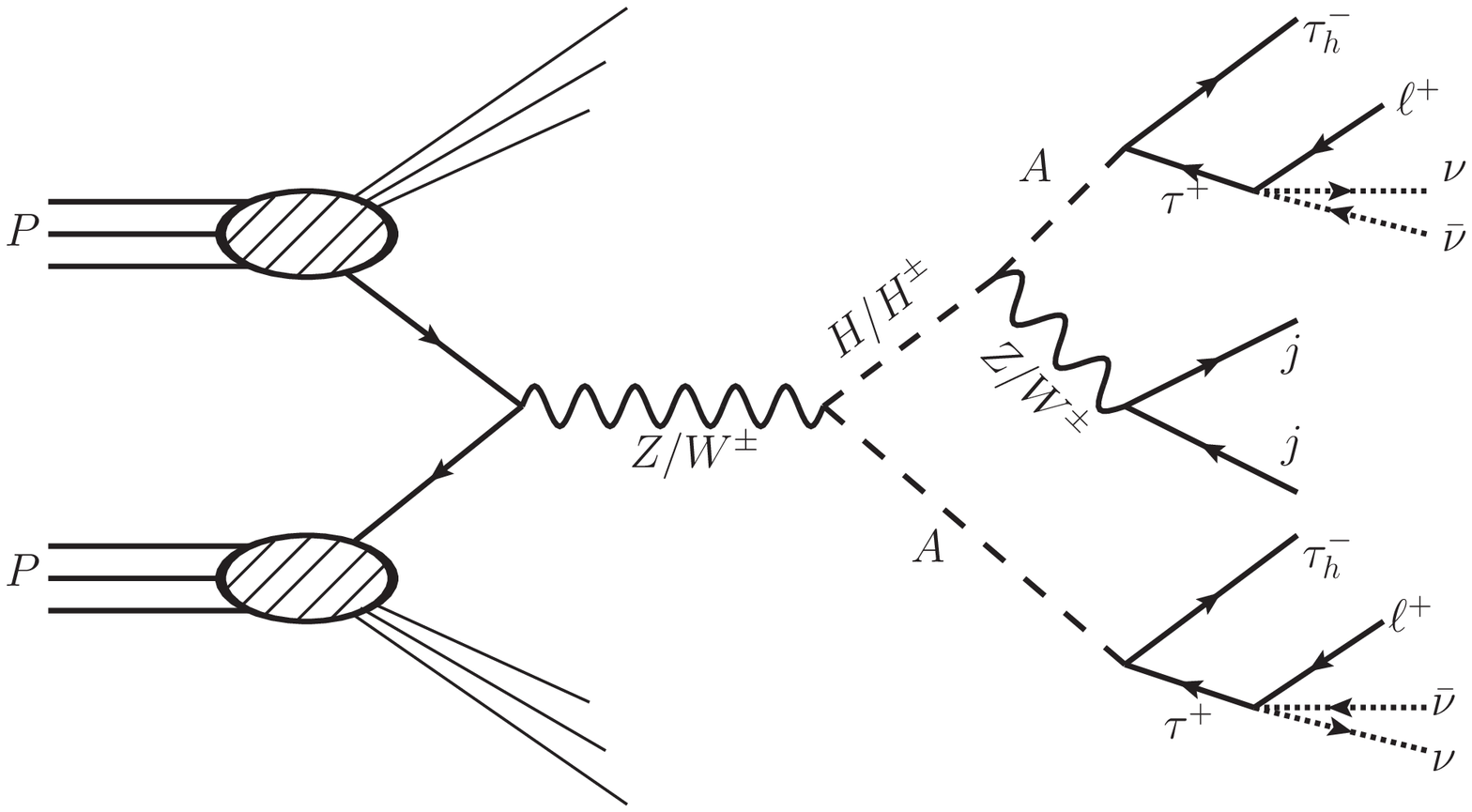}\label{subfig:feyn}}\hfill
\subfloat[]{\includegraphics[width=0.38\textwidth]{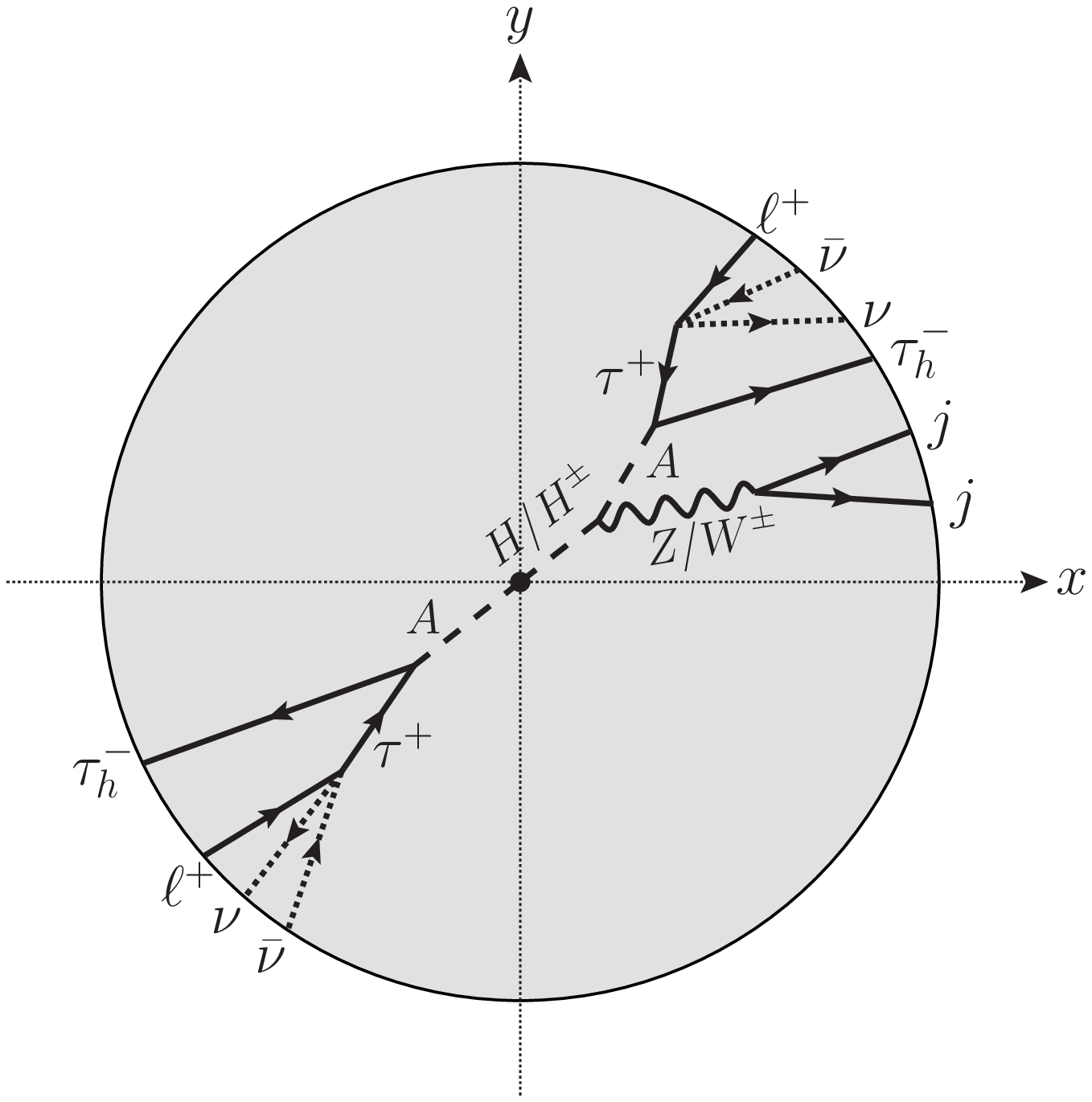}\label{subfig:feyn2}}
\caption{(a)
Representative Feynman diagram for the signal process $pp\to HA$ and
$pp\to H^\pm A$ along with the final states after the subsequent
cascade decays. (b) Representative schematic diagram of different
objects projected in the $x$-$y$ plane of the collision.}
\end{center}
\end{figure}

We, therefore, propose a subset of the 4$\tau$ final state, in which SM backgrounds
can be managed better. In order to do
so, we make use of the leptonic decay modes of two of the four
$\tau$'s. Although the branching ratio in this channel is modest, the
cleanliness of the lepton detection compensates for its low branching
ratio. More precisely, we look for those events where the two leptons
for $\tau$-decays are of the same signs. At the same time, two
$\tau$-jets of the same sign are tagged, thanks to the high
$\tau$-jets charge identification efficiencies in the one-and
three-prong channels, as already mentioned\,\cite{CMS:2022prd}. Thus
the final state we look at is ($2\ell^\pm+2\tau_h^\mp+$jets). A
representative Feynman diagram of our signal cascading all the way down
to the final signal is
illustrated in Figure~\ref{subfig:feyn}. The SM backgrounds are
substantially reduced on demanding the same charges for the
$\tau$-jet pair and at the same time for the lepton-pair.

%=====================================================================
%=====================================================================
The main backgrounds considered in this analysis are
\begin{eqnarray}
pp& \to& VV + \text{jets}\\
&&\hookrightarrow 2\tau^\pm + 2\tau^\mp + \text{jets} \nonumber\\
&&\hookrightarrow \tau^+\tau^- \ell^\pm \nu_\ell + \text{jets}, \nonumber\\
pp& \to& t\bar t,\\
pp& \to& t\bar t V,\\
pp& \to& \tau^+ \tau^- + \text{jets}.
\end{eqnarray}

The primary background for the signal comes from $VV+$jets, where
$V=\gamma^*,Z,W^\pm$. In the case of both of the vector bosons being
$\gamma^*$ and $Z$, it can directly produce
$2\tau^\pm+2\tau^\mp+$jets, and thereby end up becoming an irreducible
background. On the other hand, if one vector boson is $W^\pm$, which
decays to leptons and the other $V$ being $Z$ or $\gamma^*$ decays to
$\tau^\pm$ pairs, it can also give rise to two same-sign leptons. In that
case, one lepton comes from a $W$ boson and the other comes from a
$\tau$. Two same-sign $\tau$s do not come directly. However, a QCD jet
mistagged as a $\tau_h$ gives rise to two same-sign $\tau_h$s in an
event. In our study, we have generated the $VV+$jets background events
in the two above-mentioned scenarios explicitly, i.e.
$pp\to4\tau+$jets and $pp\to 2\tau+W+$jets. Another set of important
background channels turns out to be $t\bar t$ and
$t\bar t V$ because of their large
cross sections. One of the same-sign leptons appears from the
semileptonic decay of $B$ meson and the other appears directly from
the $W^\pm$ decays.

For the backgrounds, the parton-level events were generated at the
leading order (LO) in QCD and QED coupling. Then an appropriate
$k$-factor has been multiplied with the cross section in each of the
backgrounds to make up for the correction at the next-to-leading-order
(NLO) for $t\bar t V$ background, and at the
next-to-next-to-leading-order (NNLO) for $VV+$jets and $t\bar t$
backgrounds. The $k$-factors are 1.38, 1.57, 1.60, and 2.01, 1.72 for
$t\bar t Z$\, \cite{Kardos:2011na}, $t\bar t W$\, \cite{vonBuddenbrock:2020ter}, $t\bar t$ \, \cite{Czakon:2013goa},
$\tau^+\tau^- W + \text{jets}$ \, \cite{Grazzini:2016swo},
and $2\tau^\pm + 2\tau^\mp + \text{jets}$ \,\cite{Cascioli:2014yka}, respectively. 

%=====================================================================
\begin{figure}[!h]
\subfloat[]{\includegraphics[width=0.48\textwidth]{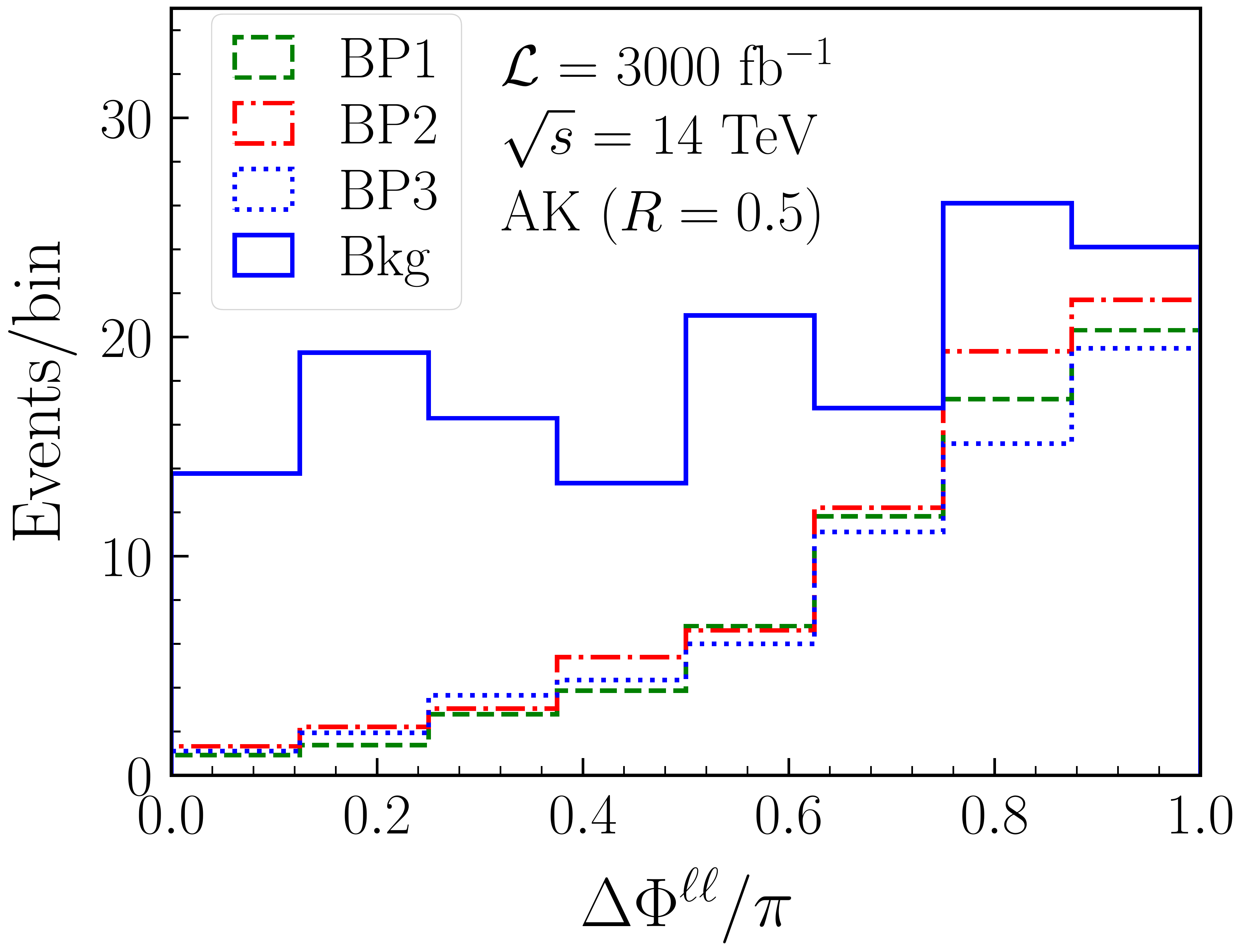}\label{fig:dphillAK}}\hfill
\subfloat[]{\includegraphics[width=0.48\textwidth]{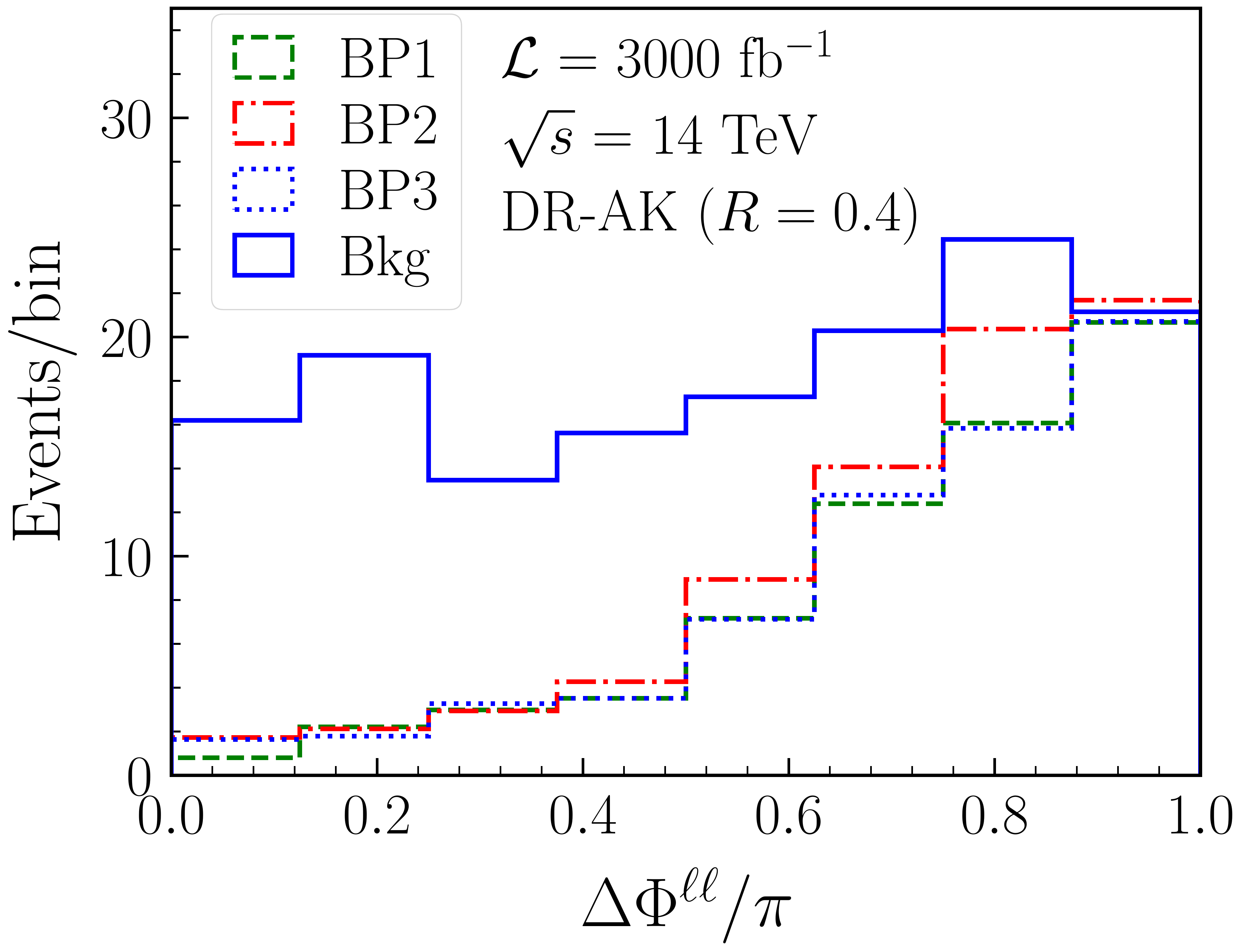}\label{fig:dphillDRAK}}\\
\subfloat[]{\includegraphics[width=0.48\textwidth]{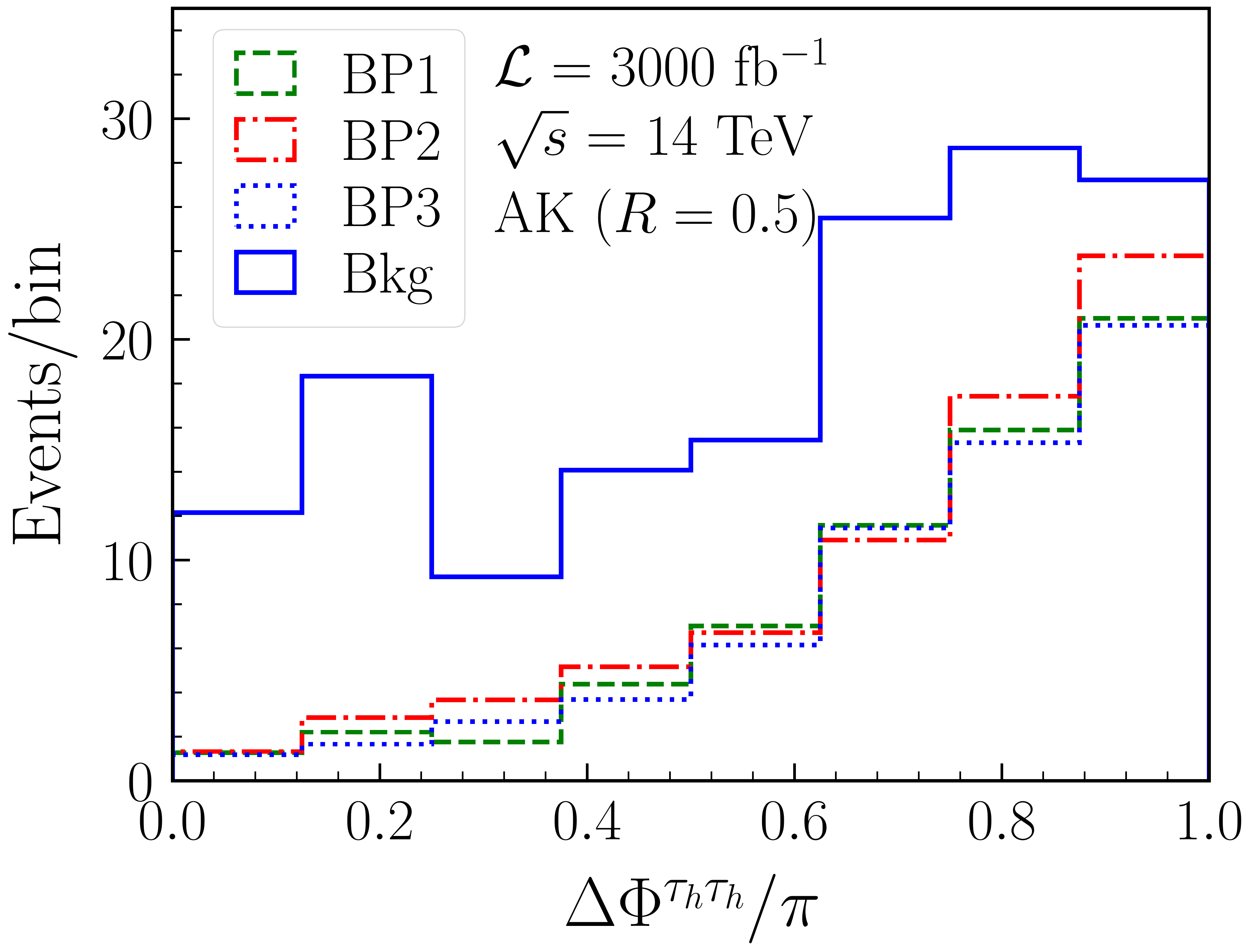}\label{fig:dphitataAK}}\hfill
\subfloat[]{\includegraphics[width=0.48\textwidth]{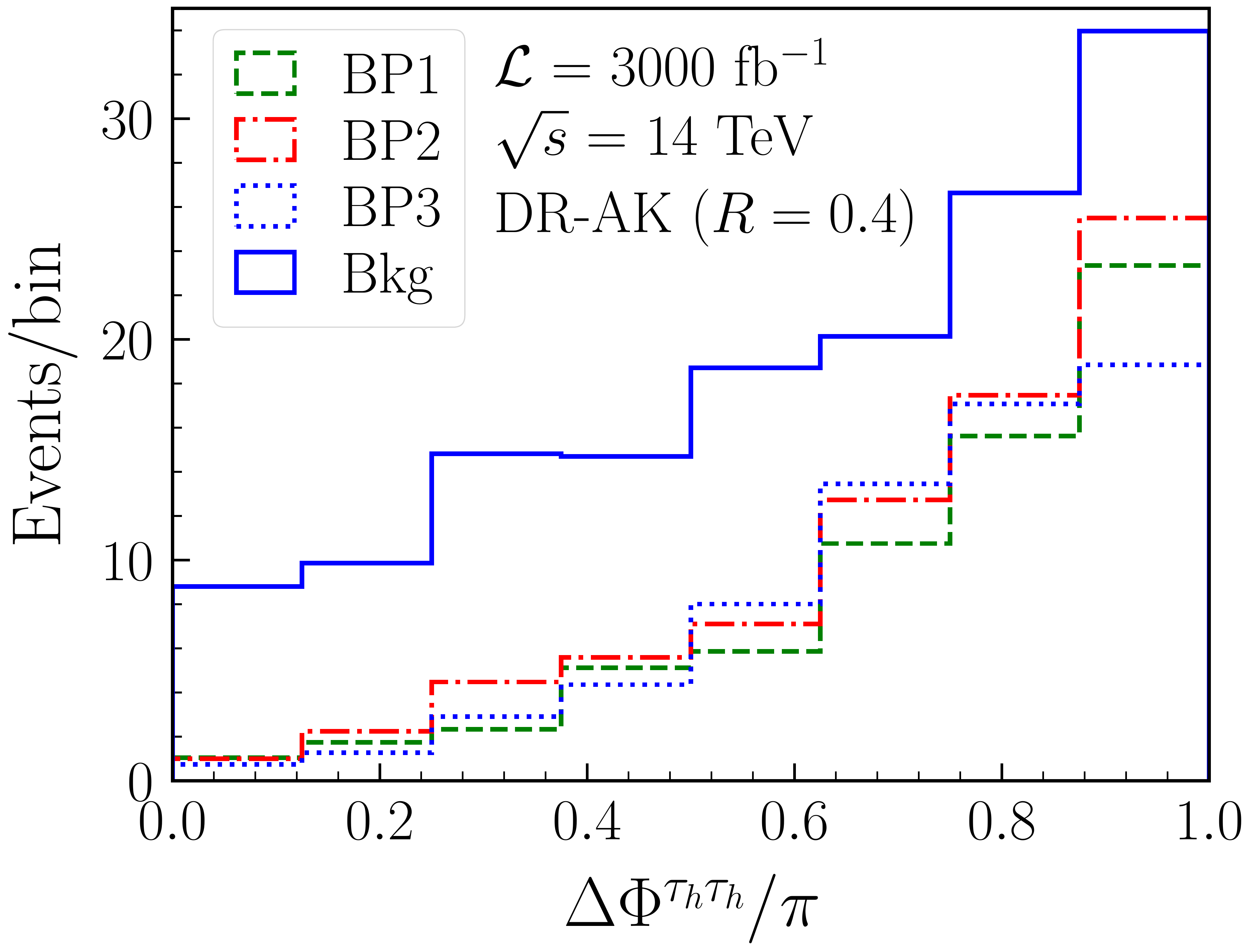}\label{fig:dphitataDRAK}}\\
\subfloat[]{\includegraphics[width=0.48\textwidth]{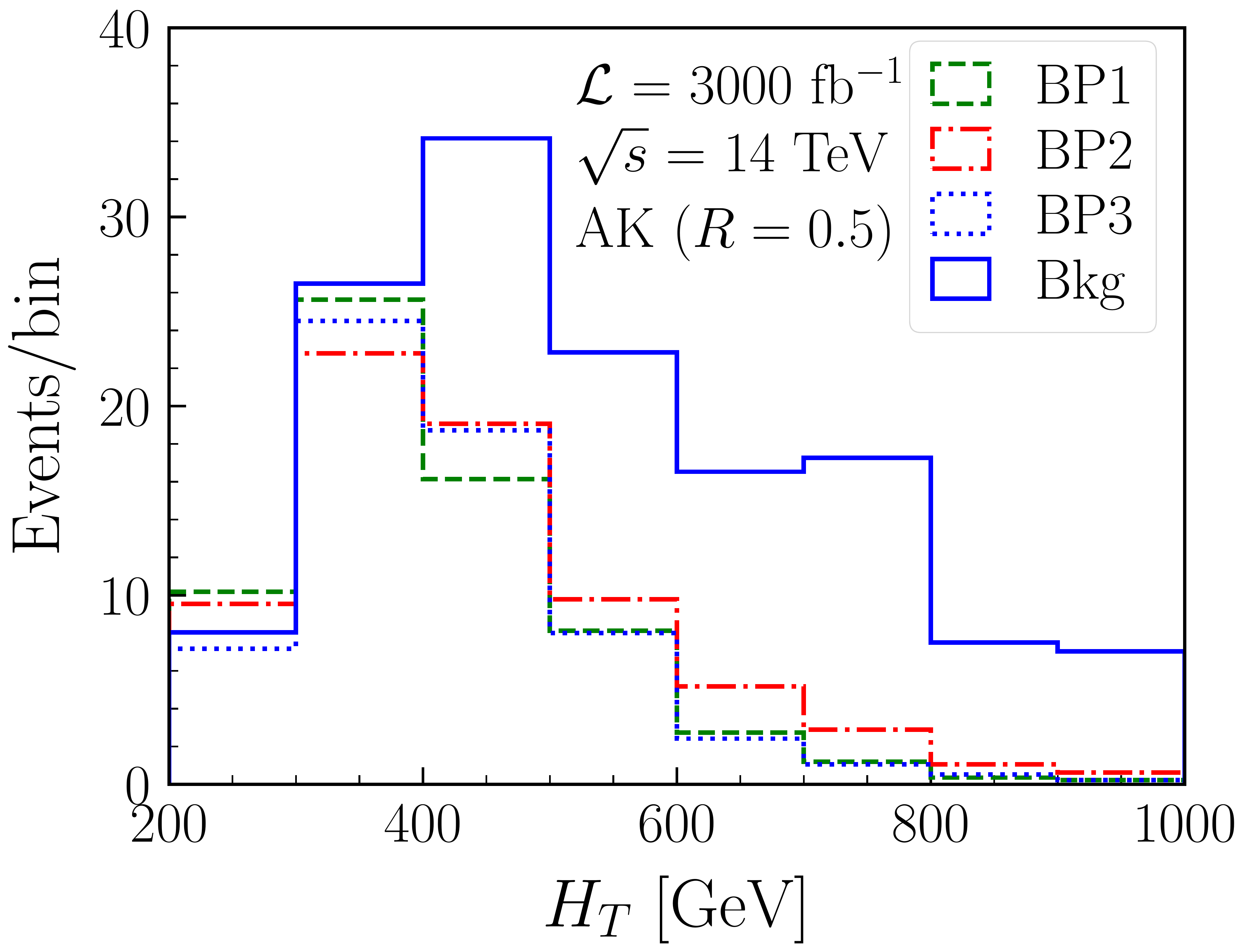}\label{fig:htAK}}\hfill
\subfloat[]{\includegraphics[width=0.48\textwidth]{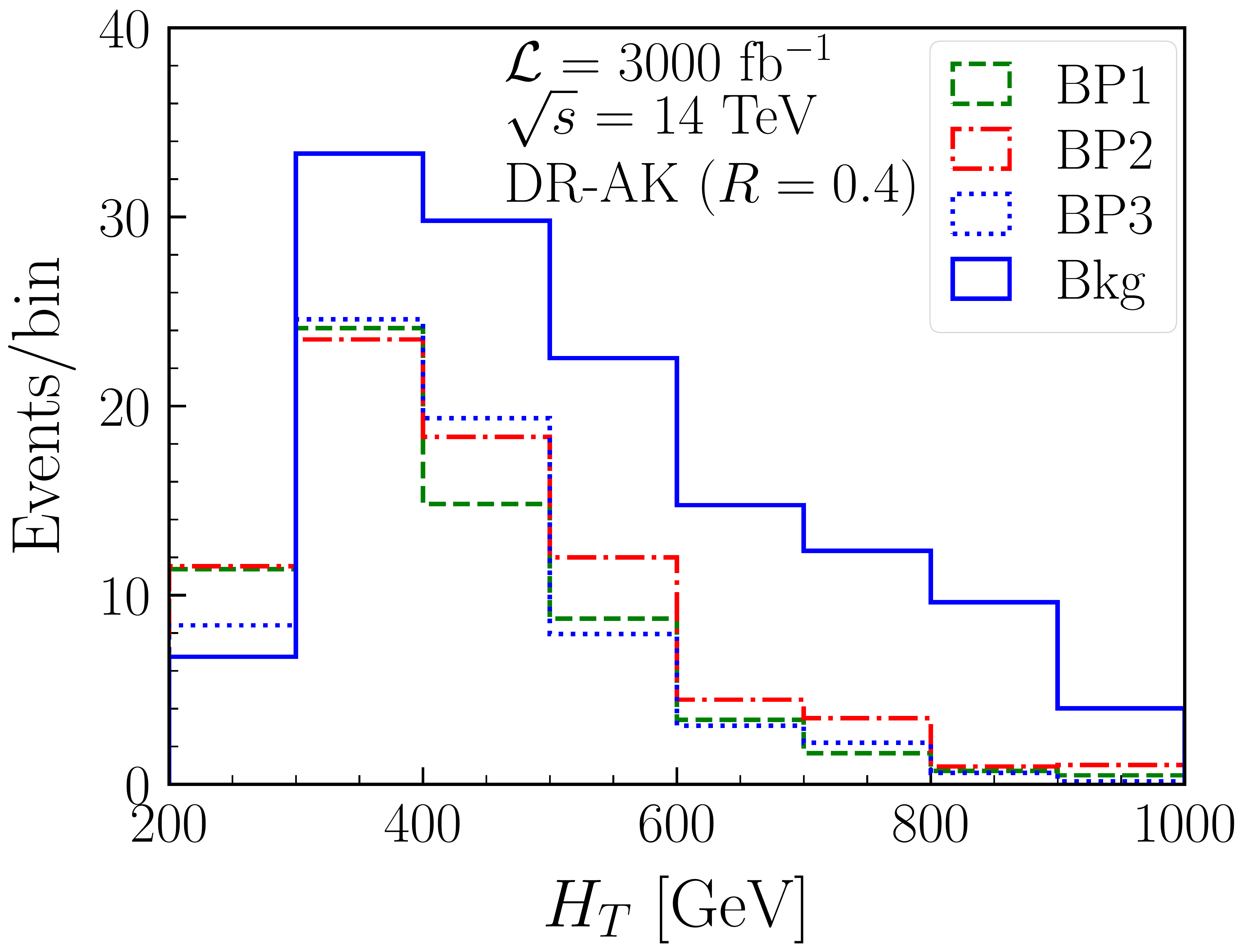}\label{fig:htDRAK}}
\caption{Distribution in different event variables for signals and backgrounds.
The left column is the distribution when the jets are clustered using
anti-$k_t$ algorithm with radius $R=0.5$. The right column is for the dynamic
radius anti-$k_t$ algorithm with initial radius $R=0.4$.}
\label{fig:dist}
\end{figure}
%=====================================================================
Before entering into further discussion on the background reduction
strategy, we outline our tools and analysis procedure. The
parton-level signal and the background events have been generated
using {\tt Madgraph5}\,\cite{Alwall:2014hca}. These parton-level
events have then been showered and hadronized by the
{\tt Pythia8}\,\cite{Sjostrand:2014zea} event generator. We used the
fast detector simulator {\tt Delphes}\,\cite{deFavereau:2013fsa} for
the simulation of detector effects. We employed two
separate algorithms for the formation of jets from the delphes
{\tt eflow} output\footnote{Operationally, the delphes-implemented
{\tt eflow} outputs are closer to Particle Flow output at the
CMS\,\cite{deFavereau:2013fsa}.}: (a) standard anti-$k_t$
algorithm (AK)\,\cite{Cacciari:2008gp} with radius 0.5, and (b)
dynamic radius anti-$k_t$ (DR-AK)
algorithm\,\cite{Mukhopadhyaya:2023rsb} recently developed by some of us with an initial radius 0.4.
For the tagging of $\tau$-jets, we used the delphes $\tau$-tagger with
efficiency 0.6 and misidentification efficiency $0.01$. 

Since the signal is primarily coming from the cascade decay of two
particles with masses in the range 60-200~GeV produced in hard
scattering, the two same-sign leptons, as well as the two same-sign
$\tau_h$'s, tend to have large separations in the azimuthal plane as
illustrated schematically in Figure~\ref{subfig:feyn2}. This is
exhibited in Figures~\ref{fig:dphillAK} and \ref{fig:dphillDRAK} via the
distribution of $\Delta\phi(\ell_1,\ell_2)$, where $\ell_1$ and $\ell_2$ are
the $p_T$-wise leading and subleading leptons, respectively. For both
cases, the signals have peaks at $\Delta\phi=\pi$ representative of
the mentioned feature for the signal. The combined background is more
of a uniform distribution in this variable. A similar feature is seen
in Figures~\ref{fig:dphitataAK} and \ref{fig:dphitataDRAK} for the $\Delta\phi$
between the two same-sign $\tau$-jets. Another important variable
$H_T$, defined as
\begin{eqnarray}
H_T = \sum_{i\in \text{visible}} \left|\vec{p}_T^{~i}\right|,
\end{eqnarray}
is particularly useful in discriminating signals from the background.
The distribution of this variable is plotted in Figures~\ref{fig:htAK}
and \ref{fig:htDRAK}. In both AK and DR-AK cases, we can see
that the background tends to have higher $H_T$ compared to that of
the signals. So, overall, a lower cut on the $\Delta\phi$ between the
two same-sign leptons or the two same-sign $\tau$-jets and an upper
cut on the variable $H_T$ are useful in separating signals from the
backgrounds. 

%%%%%%%%%%%%%%%%%%%%%%%%%%%%%%%%%%%%%%%%%%%%%%%%%%%%%%%%%%%%%%%%%%%%%%
\subsection{Result and Discussions} \label{sec:res}
We are now ready to present signal vs.~background analyses. For the
current study, we performed cut-based analyses with an integrated
luminosity of $\mathcal{L} = 3000$~fb$^{-1}$. We employed the
following set of kinematic acceptance cuts on the delphes-generated
same-sign leptons, same-sign $\tau$-jets, and QCD jets.
\begin{eqnarray}
\text{Acceptance Cuts:} \quad
   \left\{\ \begin{matrix}
   p_T^{\ell_1} > 20~\text{GeV}, && p_T^{\ell_2} > 15~\text{GeV},
&& |\eta_\ell| < 2.5, \\
   p_T^j > 30~\text{GeV}, && p_T^{\tau_h} > 30~\text{GeV},
&& |\eta_{j,\tau_h}| < 4.7, \\
   \!\!\!\!\!\!\!\!\!\!\!\!\!\!\!\!N_{\ell^\pm} = 2, &&\!\!\!\!\!\!\!\!\!\!\!\!\!\!\!\!\!\!\!\!\!\!\!\!\!\!\!\!\!\!\!\!\!\!\!\!\!\!\!\!\! N_{\tau_h^\mp} = 2, &&\!\!\!\!\!\!\!\!\!\!\!\!\!\!\!\!\!\!\!\!\!\!\!\!\!\!\!\!\!\!\!\!\!\!\!\!\!\!\!\!\!\!\!\!\!\!\!\!\!\!\!\!\!\!\!\!\!\!\!\!\!\!\!\! N_j \ge 2, &&\!\!\!\!\!\!\!\!\!\!\!\!\!\!\!\!\!\!\!\!\!\!\!\!\!\!\!\!\!\!\!\!\!\!\! \Delta R(\tau_h,\tau_h) > 0.6.
   \end{matrix} \right. \label{eqn:acc}
\end{eqnarray}
As mentioned previously, we tried two different methods, namely AK and
DR-AK algorithms, to cluster the jets from the delphes {\tt eflow}
outputs. These two algorithms show very similar distributions for the
signal and for the backgrounds as shown in Figure~\ref{fig:dist}.

We have applied lower cuts (as selection cuts) on the $\Delta\phi$
between the two same-sign leptons and between the two same-sign
$\tau$-jets. In Table~\ref{tab:cutflow}, we list the details of
the cuts and the number of events accepted after the specified cuts
for the three benchmark signals and for the backgrounds. As expected
from the distribution in $\Delta\phi$ (shown in
Figure~\ref{fig:dist}), the background is reduced by a factor of two
whereas the signals are reduced only by 20\,\%. Additionally, the cut on
the variable $H_T$ further reduces the total background by another
factor of two with less than 10\,\% reduction in the signals.  
%=====================================================================
\begin{table}[!h]\renewcommand\arraystretch{1.25}
\begin{center}
\begin{tabular}{|c|c|c|c|c|c|c|c|c|}
\hline  & \multicolumn{8}{c|}{Number of events at $\mathcal{L}=3000$~fb$^{-1}$ at $\sqrt{s}=14$~TeV LHC}\\
\cline{2-9}
  
Cuts & \multicolumn{2}{c|}{\quad BP1 \qquad}
& \multicolumn{2}{c|}{\quad BP2 \qquad}
& \multicolumn{2}{c|}{\quad BP3 \qquad}
& \multicolumn{2}{c|}{Backgrounds} \\
\cline{2-9}
& AK & DR-AK & AK & DR-AK & AK & DR-AK & AK & DR-AK \\
\hline
Acceptance (Eq.~\ref{eqn:acc}) & 65  & 66 & 62 & 66 & 72 & 76 & 138 & 137\\
\hline
$\Delta\phi(\ell_1,\ell_2) \geq 1.5$ & 57 & 56  & 53 & 57 & 60 & 66 & 82 & 78\\
\hline
$\Delta\phi(\tau_1,\tau_2) \geq 1.0$ & 53 & 53 & 49 & 54 & 55 & 61 & 68 & 66\\
\hline
$H_T \leq 500$ & 43 & 41 & 40 & 43 & 40 & 44 & 34 & 37\\
\hline
\end{tabular}
\caption{The cut-flow table for the $2\ell^\pm+2\tau^\mp+$jets channel.
The number of events after the specified cuts are shown for standard
anti-$k_t$ (AK) with radius $R=0.5$, and dynamic radius anti-$k_t$
(DR-AK) algorithm with initial radius $R=0.4$ for the signals and
backgrounds.}
\label{tab:cutflow}
\end{center}
\end{table}
%=====================================================================

The collider experiments, in general, are susceptible to systematics
uncertainty. In the HL-LHC also, we expect a certain amount of
uncertainty. We, therefore, choose to present the signal significance
with systematic uncertainties
\begin{eqnarray}
\mathfrak{S} 
  =\sqrt{2}\left[(S+B)\ln\left(1+\frac{S}{B+\epsilon^2 B(S+B)}\right)
  -\epsilon^{-2}\ln\left(1+\frac{\epsilon^2 S}{1+\epsilon^2 B}\right)
   \right]^{\frac{1}{2}},
\end{eqnarray}
where $B (S)$ are the number of background (signal) events after the
selection cuts at a given luminosity and $\epsilon$ is the overall
systematic uncertainty fraction. We tabulate the signal significance for
the three benchmark points in Table~\ref{tab:signi} for four selected
systematic uncertainties (5\,\%, 10\,\%, 15\,\%, 20\,\%). The expected
signal significance for all the benchmark points is quite good. For
all the benchmark points, the significances are approximately 5$\sigma$ with
10\,\% systematics and are well above 3$\sigma$ even with 20\,\%
systematics, which is moderately high according to the current run of
the LHC. The two methods, namely the fixed radius and the dynamic
radius anti-$k_t$ algorithm, of forming jets yield almost similar
results indicative of the performance of the DR-AK algorithm at per with
the traditional AK algorithms. 

%=====================================================================
\begin{table}[!h]\renewcommand\arraystretch{1.25}
\begin{center}
\begin{tabular}{|c|c|c|c|c|c|c|}
\hline
& \multicolumn{6}{c|}{Significance ($\mathfrak{S}$)} \\
\cline{2-7}
  \quad Systematics \qquad
& \multicolumn{2}{c|}{BP1}
& \multicolumn{2}{c|}{BP2}
& \multicolumn{2}{c|}{BP3} \\
\cline{2-7}
& \quad AK ~~\quad & DR-AK & \quad AK ~~\quad & DR-AK & \quad AK ~~\quad &  DR-AK  \\
\hline
$ 5\,\%$ & 6.0 & 5.5 & 5.6 & 5.7 & 5.6 & 5.8 \\
\hline
$10\,\%$ & 5.2 & 4.8 & 4.9 & 4.9 & 4.9 & 5.0 \\
\hline
$15\,\%$ & 4.5 & 4.0 & 4.1 & 4.2 & 4.2 & 4.2 \\
\hline
$20\,\%$ & 3.8 & 3.4 & 3.5 & 3.5 & 3.5 & 3.6 \\
\hline
\end{tabular}
\caption{The signal significances at the $\sqrt{s} = 14$~TeV LHC at
$\mathcal{L} = 3000$~fb$^{-1}$, for different levels of systematics.
The significances are shown for standard anti-$k_t$ (AK) with radius
$R=0.5$, and dynamic radius anti-$k_t$ (DR-AK) algorithm with initial
radius $R=0.4$ for the signals and backgrounds.}
\label{tab:signi}
\end{center}
\end{table}
%=====================================================================

While the AK algorithm has been in use for quite some time, the DR-AK
scheme\,\cite{Mukhopadhyaya:2023rsb}, recently developed by us, has
been profitably used in other contexts, especially when the physical
origin of jets of differing radii are to be distinguished. As can be
seen from Figure~\ref{fig:dist} and Tables~\ref{tab:cutflow} and
\ref{tab:signi}, the new algorithm is competitive and in fact performs
better in some kinematic regions. 

The result presented here is for 3000~fb$^{-1}$ integrated luminosity,
for which the signal significance often rises to the discovery level in the
$2\ell^\pm+2\tau^\mp+$jets channel.
However because of its clean nature, the signal
starts having significance exceeding 3$\sigma$ even at 1000~fb$^{-1}$,
provided the systematics can be brought under sufficient control
(within 10\%). This is indeed a possibility in the CMS phase-2
detector at the HL-LHC with the improved detector sensitivity in the
CMS detectors\,\cite{Contardo:2015bmq}. Furthermore, with the improved $\tau$-tagging
efficiency due to the incorporation of an online L1 tracker
trigger\,\cite{Zabi:2020gjd} at the CMS, finding the signal in our proposed channel
can indeed be of high significance.

%%%%%%%%%%%%%%%%%%%%%%%%%%%%%%%%%%%%%%%%%%%%%%%%%%%%%%%%%%%%%%%%%%%%%%
\section{Summary and Conclusion} \label{sec:summ}
Type-X 2HDM is a phenomenologically well-motivated BSM scenario.
We have performed a scan over the parameter space of this model, taking
into account the constraints coming from the theoretical consistency,
measurement of the electroweak oblique parameter at the LEP, and
various scalar searches at the LHC. The scalar searches at the LHC
constrain the mass of the CP-odd scalar ($m_A$) to be above $m_h/2$
primarily because of the non-observation of any significant anomaly in
the $h\to4\tau$ channel. On the other hand, the electroweak oblique
parameter measurements prefer a region where $m_H \simeq m_{H^\pm}$.
The anomalous magnetic moment of the muon prefers low $m_A$ and
relatively high $\tan\beta$ regions.

We have chosen three benchmark points consistent with the constraints
discussed above to examine clean collider signatures in the channel
having two same-sign leptons, two same-sign $\tau$-jets, and at least
two jets at the HL-LHC. With the fixed-radius anti-$k_t$ algorithm, we
achieve approximately 5$\sigma$ signal significance with moderate
systematics of $10\,\%$ at 3000~fb$^{-1}$ integrated luminosity. A
conservative scenario with 20\,\% systematics is also able to
yield more than 3$\sigma$ signal significance. We have parallelly
performed the analysis using recently proposed dynamic radius
jet clustering algorithm, which produces similar results as the
traditional anti-$k_t$ algorithm and thereby establishing the validity
of the proposed algorithm. We further note that the signal is likely
to appear even with  1000~fb$^{-1}$ luminosity at the CMS phase-2
detector with the projected improved sensitivity in the tracker and
enhanced efficiency in the $\tau$-tagging\,\cite{Contardo:2015bmq,Zabi:2020gjd}.

%%%%%%%%%%%%%%%%%%%%%%%%%%%%%%%%%%%%%%%%%%%%%%%%%%%%%%%%%%%%%%%%%%%%%%
\section*{Acknowledgements}
The authors thank Atri Dey for useful discussions.
The authors acknowledge the support of the Kepler Computing facility
maintained by the Department of Physical Sciences, IISER Kolkata for
various computation needs. S.S. thanks the Council of Scientific
\& Industrial Research (CSIR), Government of India for financial
support.

%%%%%%%%%%%%%%%%%%%%%%%%%%%%%%%%%%%%%%%%%%%%%%%%%%%%%%%%%%%%%%%%%%%%%%
% \clearpage
\bibliographystyle{JHEP}

\providecommand{\href}[2]{#2}\begingroup\raggedright\endgroup

\end{document}